\documentclass[12pt]{article}
\usepackage{amsbsy,amsmath}
\usepackage{amsfonts}
\usepackage{amssymb}
\usepackage{amscd}
\usepackage{bbm}
\usepackage{fancybox}
\usepackage{cite}
\usepackage{amsmath,amsfonts,amsbsy}
\usepackage{pstricks,pst-node}
\usepackage[small,bf,hang]{caption2}
\usepackage{graphicx}
\usepackage{epsfig}
\usepackage{psfrag}
\usepackage{comment}

\usepackage{float}

\psset{unit=1.3cm,linewidth=.5pt,radius=.2}  

\usepackage{multirow}                     
\usepackage{float}                          
\usepackage{lscape}                         
\usepackage{bm}

\newcommand{\bb}{\bar\beta}

\newcommand{\beq}{\begin{equation}}
\newcommand{\eeq}{\end{equation}}
\newcommand{\bi}{\begin{itemize}}
\newcommand{\ei}{\end{itemize}}
\newcommand{\bt}{\begin{tabular}}
\newcommand{\et}{\end{tabular}}
\newcommand{\bc}{\begin{center}}
\newcommand{\ec}{\end{center}}

\newcommand{\be}{\begin{equation}}
\newcommand{\ee}{\end{equation}}
\newcommand{\bea}{\begin{eqnarray}}
\newcommand{\eea}{\end{eqnarray}}
\newcommand{\ba}{\begin{array}}
\newcommand{\ea}{\end{array}}

\def\bbox{{\,\lower0.9pt\vbox{\hrule \hbox{\vrule height 0.2 cm
\hskip 0.2 cm \vrule height 0.2 cm}\hrule}\,}}
\newcommand{\dsl}{\pa \kern-0.5em /}

\font\mybb=msbm10 at 12pt
\def\bb#1{\hbox{\mybb#1}}

\def\bR {\bb{R}}
\def\bE {\bb{E}}
\def\bB {\bb{B}}
\def\bT {\bb{T}}

\def\bN {\bb{N}}

\def\bG {\bb{G}}
\def\bF {\bb{F}}
\def\bC {\bb{C}}

\def\bK {\bb{K}}
\def\bC {\bb{C}}
\def\bD {\bb{D}}
\def\bA {\bb{A}}

\def\bfo{\mbox{\boldmath $\omega$}}

\def\bfn{\mbox{\boldmath $\nabla$}}
\def\bfs{\mbox{\boldmath $\sigma$}}

\newcommand{\bs}[1]{\boldsymbol{#1}}
\newcommand{\PP}{\bs{\mathcal{P}}}

\makeatletter \@addtoreset{equation}{section} \makeatother

\def\slashchar#1{\setbox0=\hbox{$#1$}           
   \dimen0=\wd0                                 
   \setbox1=\hbox{/} \dimen1=\wd1               
   \ifdim\dimen0>\dimen1                        
      \rlap{\hbox to \dimen0{\hfil/\hfil}}      
      #1                                        
   \else                                        
      \rlap{\hbox to \dimen1{\hfil$#1$\hfil}}   
      /                                         
   \fi}

\begin{document}

\begin{titlepage}
\begin{center}

\vskip 1.5cm

{\Large \bf DBI in the IR }

\vskip 1cm

{\bf Luca Mezincescu\,${}^1$ and  Paul K.~Townsend\,${}^2$} \\

\vskip 25pt

{\em $^1$  \hskip -.1truecm
\em Department of Physics, University of Miami, P.O. Box 248046, \\
Coral Gables, FL 33124, USA
 \vskip 5pt }

{email: {\tt mezincescu@miami.edu}} \\

\vskip .4truecm

{\em $^2$ \hskip -.1truecm
\em  Department of Applied Mathematics and Theoretical Physics,\\ Centre for Mathematical Sciences, University of Cambridge,\\
Wilberforce Road, Cambridge, CB3 0WA, U.K.\vskip 5pt }

{email: {\tt P.K.Townsend@damtp.cam.ac.uk}} \\

\end{center}

\vskip 0.5cm
\begin{center} {\bf ABSTRACT}\\[3ex]
\end{center}

The Infra-Red limit of a planar static D3-brane in AdS$_5\times S^5$ is a tensionless D3-brane at the  Anti-de Sitter horizon with dynamics governed by a strong-field limit 
of the Dirac-Born-Infeld action, analogous to that found from the Born-Infeld action by  Bialynicki-Birula. As in that case, the field equations are those of an interacting
4D conformal invariant field theory with an $Sl(2;\bR)$ electromagnetic duality invariance, but the D3-brane origin makes these properties manifest. 
We also find an $Sl(2;\bR)$-invariant action for these equations.

\vfill

$*$ Contribution to  {\it A Passion for Theoretical Physics: in Memory of Peter Freund}. 

\end{titlepage}
\tableofcontents
\section{Introduction}

The observation in 1980 by Peter Freund and Mark Rubin  that 11D supergravity admits an AdS$_4\times S^7$ vacuum \cite{Freund:1980xh} was largely responsible for the  1980s revival of  Kaluza-Klein (KK) theory, in part because their paper emphasised the distinction between dimensional reduction (a mathematical convenience) and  ``spontaneous compactification'' (a dynamical issue) but also because it focused attention on  Anti-de Sitter (AdS) space. 

Dirac had initiated  a study of field theory on AdS in 1963  with his discovery of  ``singleton'' irreps of the AdS$_4$ isometry group that have no Poincar\'e analog \cite{Dirac:1963ta},  and this was revived in  the early 1980s by Flato and Fronsdal \cite{Flato:1980we}; they interpreted Dirac's singletons (the ``Di'' and ``Rac'')  as degrees of freedom of a conformal field theory on the AdS boundary.  Although this idea was initially independent of the concurrent resurgence of KK theory, it soon gained support from computations of the  particle spectrum in AdS$_4$ arising from the harmonic expansion of fields on $S^7$; the supersingleton irrep appearing in this expansion \cite{Sezgin:1983ik}  is most simply interpreted in terms of an ${\cal N}=8$  superconformal field theory on the AdS$_4$ boundary \cite{Nicolai:1984gb}. 

Following the discovery  in 1987 of the 11D supermembrane \cite{Bergshoeff:1987cm} it was argued that the AdS$_4$ boundary degrees of freedom were those of a  ``membrane at the end of the universe''  \cite{Bergshoeff:1987dh}.  Following the further discovery in 1990 of what we now call the M2-brane solution of 11D supergravity \cite{Duff:1990xz},  it was pointed out  that it interpolates between the 11D Minkowski vacuum at transverse infinity and  the AdS$_4\times S^7$ vacuum in a near-horizon 
limit \cite{Gibbons:1993sv}. This confirms the connection between (super)membrane dynamics and (super)singletons because what looks like a membrane from transverse infinity looks like an AdS boundary from  the near-horizon region. 

There was, however, a problem with the idea that singletons are essentially dynamical degrees of freedom of a membrane. The standard worldvolume action for a membrane,
proposed by Dirac in 1961 \cite{Dirac:1962iy},  is  not conformal invariant. The  free-field theory of small fluctuations about the Minkowski worldvolume of a planar static membrane  is conformal invariant but this appears to be broken by the interactions.    This puzzle was largely resolved (post AdS/CFT)  in  \cite{Claus:1998mw} where it was shown (i) 
that the action for a membrane is invariant under the isometries of the spacetime background,  and (ii) that  a planar membrane in  the 11D Freund-Rubin vacuum 
can be static at any value of the radial coordinate $r$ in coordinates adapted to the foliation of AdS$_4$ by 3D Minkowski `slices';  in this context the AdS$_4$ isometry group is naturally interpreted as the 3D conformal group, spontaneously broken to the 3D  Poincar\'e group. The radial coordinate $r$, viewed as a worldvolume field, is the Nambu-Goldstone scalar corresponding to broken scale invariance; a  scale transformation shifts the value of $r$, either towards the AdS boundary at  $r=\infty$ (the UV limit) or towards a Killing horizon at $r=0$.  Conformal invariance is restored on the AdS boundary but this is also a free-field limit. 

These days there is little interest in this precursor to AdS/CFT because it is viewed as describing only the `centre-of-mass' motion of a macroscopic brane, whereas 
the focus since Maldacena's famous work \cite{Maldacena:1997re} is on the inter-brane dynamics.  However, we aim to convince the reader that there is still
an interesting unexplored corner of the older circle of ideas, which involved a single brane (a ``probe''  since it is assumed to 
have negligible back-reaction).  The spontaneously broken conformal invariance of the worldvolume field theory on a static planar membrane in AdS$_4$
is restored not only if it coincides with the AdS boundary, as mentioned above,  but also if it coincides with the Killing horizon. This is the IR limit; it is also a limit in which 
the membrane is tensionless. 

The same set of ideas apply to the  AdS$_5\times S^5$ vacuum of IIB 10D supergravity \cite{Schwarz:1983qr}; its discovery was motivated by the Freund-Rubin 
solution of 11D supergravity,  but its importance was not apparent at the time.  A decade later, after the construction of a class of supersymmetric brane solutions of 
the maximal 10D  supergravity theories \cite{Horowitz:1991cd},  it was shown to be the near-horizon limit of what would later be called the D3-brane solution of IIB  supergravity  \cite{Gibbons:1993sv}. In precise analogy with the M2-brane case, one may consider a static planar probe D3-brane in this  D3-brane background solution of IIB supergravity.  This was studied in \cite{Claus:1998mw}, and the relation to singletons was further explored in \cite{Ferrara:1997dh}, but here we investigate the nature of the worldvolume field theory on the probe D3-brane in the IR limit  for which it coincides with AdS$_5$ Killing horizon.  As for the M2 case, this limit leads to a tensionless brane, but the D3 case has some interesting additional  features. 

In a 10D Minkowski background (and for a particular choice of the constant values of the dilaton and axion fields of IIB supergravity)  the bosonic  worldvolume field theory of the D3-brane, which is all that we need here,  can be 
determined by string-theory calculations \cite{Fradkin:1985ys, Callan:1988wz, Leigh:1989jq,Li:1995pq,Douglas:1995bn}.  In a Minkowski vacuum background the action is of Dirac-Born-Infeld type:
\begin{equation}\label{BIaction}
S= -T \int d^4\xi \sqrt{ -\det \left(\bG+ F/\sqrt{T}\right)} \, , 
\end{equation}
where $\bG$ is the induced worldvolume metric and $F=dA$ is the field strength 2-form for a worldvolume abelian gauge potential 1-form $A$.  The constant $T$ is the
3-brane tension, which arises from the string-theory calculation in the form $T\propto (\alpha')^{-2}$, where $\alpha'$ is the inverse string tension.  When $\bG$ is the 4D Minkowski metric the DBI action reduces to the Born-Infeld action for non-linear electrodynamics; without the Born-Infeld field it reduces to the Dirac-type action for a relativistic 3-brane as the
$4$-volume of the worldvolume  in the induced metric $\bG$. 

In the AdS$_5\times S^5$ background, there is an additional term in the D3-brane  action due to its coupling to the 4-form field strength of IIB supergravity. This `Wess-Zumino' (WZ) term comes with a factor of $T$ and it cancels a term proportional to $T$ in the Hamiltonian arising from the DBI action after gauge fixing. This allows the $T\to\infty$ limit to be taken, which is equivalent (in this background) to a UV limit  in which a planar static D3-brane moves to the AdS boundary.  The result is simply a free field theory for an ${\cal N}=4$  Maxwell supermultiplet (without the fermions).

Here we are interested in the $T\to 0$ limit because in the same AdS$_5\times S^5$ background, the IR limit leads to zero tension, but we shall begin by investigating
the tensionless limit in a Minkowski background.  In the Born-Infeld case, where $\bG=\eta$ (the Minkowski $4$-metric), this limit is equivalent to the strong-field limit studied by Bialynicki-Birula in 1983 \cite{BialynickiBirula:1984tx}.  The limit cannot be taken  in the original Born-Infeld action; one must first pass to its phase-space  version, after which one arrives at the action 
\begin{equation}\label{BBE}
S_{BB}= \int\! dt \int \! d^3\sigma \left\{{\bf E}\cdot {\bf D} - \left|{\bf D}\times {\bf B}\right| \right\} \, , 
\end{equation}
where ${\bf E}$ and ${\bf B}$ are the electric and magnetic $3$-vector fields; they can be expressed in terms of the 
components $(A_0,{\bf A})$  of the 1-form potential $A$ in the usual way:
\begin{equation}\label{BBEact}
{\bf E} = \bfn A_0 - \dot{\bf A} \, , \qquad   {\bf B} = \bfn \times {\bf A}\, . 
\end{equation}
The $3$-vector field ${\bf D}$  is canonically conjugate to $-{\bf A}$; it is also (prior to the strong-field limit) the electric displacement  field\footnote{Within the Hamiltonian formulation of electrodynamics expounded in \cite{BialynickiBirula:1984tx}  the electric field ${\bf E}$ is defined as the variation of the Hamiltonian with respect to the electric-displacement field ${\bf D}$; in the context of the phase-space action this is just the field equation that follows from variation of ${\bf D}$.}. The phase-space action of (\ref{BBE}) defines a non-linear 4D conformally invariant theory that we shall call ``Bialynicki-Birula electrodynamics'' (BBE). One of its most interesting features, is an $Sl(2;\bR)$ electromagnetic duality invariance of its field equations \cite{BialynickiBirula:1984tx} (which generalises the standard $SO(2)$ duality invariance of the Born-Infeld field equations). 

To investigate the analogous strong-field limit of the D3-brane action we need the phase-space formulation of the DBI action for $p=3$. The phase-space action for the more general case of the super-Dp-brane,  for any $p$, including wordvolume fermions and  a generic bosonic  IIA or IIB supergravity background, is already known \cite{Bergshoeff:1998ha}, and  the bosonic part of this Dp-brane action in a Minkowski background was used in  \cite{Gauntlett:1997ss} for an investigation into ``worldvolume solitons''.  For the convenience of the reader, we present a few details of the derivation of this bosonic result, unencumbered by fermions\footnote{We 
should point out here that  a different phase-space action was obtained in \cite{Lindstrom:1997uj}, and this was used to obtain a different tensionless limit. At present, we
do not understand the relation of the results of \cite{Lindstrom:1997uj} to either those of \cite{Bergshoeff:1998ha} or those found here.}. One new result is the Poisson-bracket (PB) algebra  of the phase-space constraints; its closure is confirmation that the set of constraints is first-class, and hence that they generate gauge invariances of the action.

Once the DBI action has been replaced by its phase-space version, the zero tension limit may be taken, and this can be done for any spacetime dimension
$D=(p+1)+n$, and any spacetime metric. The result for $p=3$,  after gauge fixing, is precisely the BBE action when $n=0$; i.e. when the D3-brane is `space-filling'. 
For $n>0$ we obtain a generalization to include $n$  scalar fields. We shall see that the $n=1$ case is relevant to the IR dynamics of a D3-brane in AdS$_5$. 

Another aim of this article is to show how certain features of BBE are explained by their D3-brane origin. For example,  the BBE action (\ref{BBE}) is Lorentz invariant, but this is not manifest. From the D3-brane perspective this is because a gauge choice has been made that breaks  Lorentz invariance, implying its non-linear realization. Prior to 
gauge fixing the Lorentz invariance is manifest because the BI fields are Lorentz scalars! We explain how this paradox is resolved.  It also easier to understand why BBE is a conformally invariant  theory from its D3-brane origin. Finally, the  otherwise surprising emergence of an $Sl(2;\bR)$ electromagnetic duality invariance is easily understood from the
D3-brane perspective. We also present an alternative, but equivalent,  phase-space action that is manifestly $Sl(2;\bR)$ invariant, with corresponding Noether charges. 

We shall begin with some preliminaries on the Hamiltonian (phase-space) formulation of Dp-brane actions, after which we specialise to the D3-brane and take its tensionless limit to arrive at BBE and its generalizations to include scalar fields.  We discuss the properties of these theories at some length before returning to the D3-brane in the AdS$_5\times S^5$ IIB vacuum, where we explain the relation of its zero tension limit to  the IR limit in which the D3-brane `vacuum' coincides with the AdS Killing horizon. We conclude with a brief summary, and a few personal remarks concerning our friendship with Peter Freund.

\section{Hamiltonian DBI  preliminaries}

The maximal 10D supergravity fields, are conveniently divided into those arising from the NS-NS sector of a type II superstring
and those arising from the R-R sector. The former comprise the spacetime metric $g$ (in Einstein conformal frame),  dilaton $\phi$ and Kalb-Ramond 2-form potential $C$,
which couple to a Dp-brane through the DBI part of its action;  the latter comprise a $(p+1)$-form field and a series of 
lower-order form fields, which couple through a WZ term. The complete (low-energy) action (omitting fermions) takes the form \cite{Douglas:1995bn}
\begin{equation}\label{Sback}
S= -  \int\! d^{p+1}\xi\, \left\{ \bT \sqrt{- \det (\bG + \bF)} \right\} + S_{WZ}\, , 
\end{equation}
where 
\begin{equation}
\bT = Te^{-\phi} \, , \qquad \bF = \ell^2 F - \bC\, . 
\end{equation}
The length parameter $\ell$ is needed for dimensional reasons; the string theory calculation gives $\ell^2 \propto \alpha'$ (the inverse string tension) so 
$\ell^2 \propto 1/\sqrt{T}$ for a D3-brane. The worldvolume metric $\bG$ and the two form $\bC$ are induced from the corresponding spacetime
fields; in worldvolume coordinates $\{\xi^\mu; \mu=0, 1, \dots, p\}$ they are
\begin{equation}
\bG_{\mu\nu} =  \partial_\mu X^m\partial_\nu X^n g_{mn}\, , \qquad \bC_{\mu\nu} = \partial_\mu X^m\partial_\nu X^n C_{mn}\, , 
\end{equation}
where $g_{mn}(X)$ and $C_{mn}$ are the components of $g$ and $C$  in spacetime coordinates $\{X^m; m=0,1, \dots, s\}$; of course, 
$s=9$ for the Dp-branes of superstring theory but what follows applies for any $s$.  The remaining Wess-Zumino term can be ignored for present purposes; it 
will just lead to redefinitions of the momentum variables that emerge from consideration of the DBI action alone. 

To pass to the phase-space form of the action we set $\xi^\mu=\{t,\sigma^i; i=1, \dots p\}$ and then define $\bK(t)$ to be the $p$-metric induced on the
$p$-brane at any given time $t$; i.e. $\bK_{ij}(t, \bfs) :=  \bG_{ij}$; in this notation we may use $\bK^{ij}$ to denote the components of the inverse $p$-metric 
$\bK^{-1}$. We also write the components of $\bF$ as 
\begin{equation}
\bE_i(t, \bfs) := \bF_{i0} \, , \qquad \bB_{ij}(t,\bfs) :=\bF_{ij}(\xi)\, .
\end{equation}
To implement a time-space split in the DBI action we use the identity \cite{Bergshoeff:1998ha}
\begin{equation}\label{BTid}
\det (\bG+ \bF) \equiv \left\{\bG_{00} - \left(\bG_{0i} - \bE_i\right) \left[\left(\bK+ \bB\right)^{-1}\right]^{ij}\left(\bG_{0j} 
+\bE_j\right)\right\}\det\left(\bK+\bB\right) \, . 
\end{equation}
A further useful identity is 
\begin{equation}
\left(\bK+\bB\right)^{-1} \equiv  \tilde \bK^{-1} -  \bK^{-1}\bB\tilde \bK^{-1}\, , \qquad \tilde \bK := \bK - \bB \bK^{-1}\bB\, . 
\end{equation}
Notice that $\tilde \bK$ is symmetric, and hence so is $\tilde \bK^{-1}$, whereas: 
\begin{itemize}

\item {\bf Lemma}: $\bK^{-1}\bB\tilde \bK^{-1}$ is antisymmetric. To prove this we use
\begin{equation}
\tilde \bK^{-1}  = \left[ 1 - (\bK^{-1}\bB) ^2 \right]^{-1} \bK^{-1}
\end{equation} 
to show that 
\begin{equation}
(\bK^{-1}\bB) \tilde \bK^{-1} = \left[1-(\bK^{-1}\bB)^2 \right]^{-1} \left(\bK^{-1}\bB \bK^{-1}\right) = \tilde \bK^{-1} (\bB \bK^{-1}) 
\end{equation}
and hence that 
\begin{equation}
\left[\bK^{-1}\bB\tilde \bK^{-1}\right]^T = - \tilde \bK^{-1}(\bB \bK^{-1})  = - \left(\bK^{-1}\bB\right)\tilde \bK^{-1} = - \bK^{-1}\bB\tilde \bK^{-1}\, . 
\end{equation}

\end{itemize}
Using these (anti)symmetry properties we deduce that 
\begin{equation}\label{simpledet}
\det (\bG+ \bF) =  \left\{\bG_{00} + \bE_i \bK^{ij} \bE_j -  \bN_i [\tilde \bK^{-1}]^{ij} \bN_j \right\}\det(\bK+\bB)\, , 
\end{equation}
where 
\begin{equation}
\bN_i := \bG_{0i} + \bE_j \bK^{jk}\bB_{ki}  \, . 
\end{equation}

We now claim that an equivalent action to (\ref{Sback}), but without the WZ term,  is 
\begin{equation}\label{Sback2}
S= \int \! dt \int \! d^p\! \sigma \, \left\{ \dot X^m P_m + \bE_i \bD^i  - e{\cal H}_0 - u^i {\cal H}_i\right\}\, , 
\end{equation}
where $e$ and $u^i$ are Lagrange multipliers for phase-space constraints with constraint functions 
\begin{eqnarray}\label{hamcons}
{\cal H}_0 &=&  \frac{1}{2} \left[ P^2 + \bD^i\bD^j \bK_{ij} + \bT^2 \det(\bK+ \bB)\right] \, , \nonumber \\
{\cal H}_i &=& \partial_i X^m P_m - \bB_{ij}\bD^j\, , 
\end{eqnarray}
where $P^2 = g^{mn}P_mP_n$.  This result is easily checked by a process of sequential elimination of variables:
\begin{itemize}

\item First eliminate $P$ and $\bD^i$ by their equations of motion
\begin{equation}
P =  e^{-1}\left( \dot X - u^i\partial_i X\right)   \, , \qquad  \bD^i = e^{-1}\bK^{ij} \left({\bE}_j - \bB_{jk}u^k\right) \, . 
\end{equation}
This yields the new action
\begin{eqnarray}
S &=& {1\over2}\int \! dt \int\!  d^p\! \sigma \Big\{e^{-1}\left[ \bG_{00} + \bE_i \bK^{ij} \bE_j \right]  
-e\,\bT^2\det\left(\bK+\bB\right) \nonumber \\ 
&& \qquad \qquad \qquad \qquad +\,  e^{-1}\left[ u^i \tilde \bK_{ij}  u^j -2u^i \bN_i \right]\Big\}\, . 
\end{eqnarray}
The p-brane vector field $u^i$ is now auxiliary and can be trivially eliminated; this yields the action
\begin{equation}
S = {1\over2} \int \! dt \int\!  d^p\! \sigma \left\{e^{-1}\left[ g_{00} + \bE_i\bK^{ij}\bE_j  
- \bN_i  [\tilde \bK^{-1}]^{ij} \bN_j \right]  - e\, \bT^2\det\left(\bK+\bB\right) \right\}\, ,
\end{equation}
Now eliminate $e$ and use (\ref{BTid}) to recover (\ref{Sback}) (without the WZ term). 
\end{itemize}
Returning now to (\ref{Sback2}), we set
\begin{equation}
\bD^i= \ell^{-2} D^i\, , 
\end{equation}
so that
\begin{equation}
\bE_i \bD^i = -\dot A_i D^i - A_0 \partial_i D^i - \ell^{-2} \bC_{i0}D^i + \partial_i\left(A_0 D^i\right)\, . 
\end{equation}
This allows us to rewrite (\ref{Sback2}) as
\begin{equation}
S= \int\! dt\int\! d^p\!\sigma \left\{ \dot X^m \left[P_m + \ell^{-2} D^i(\partial_i X^n) C_{mn} \right] - \dot A_i D^i - A_0 {\cal J} - e{\cal H}_0 - u^i {\cal H}_i \right\} \, , 
\end{equation}
where $A_0$ is now a Lagrange multiplier for the constraint that generates the abelian gauge transformation of $A_i$; the new constraint function is
\begin{equation}
{\cal J} = \partial_i D^i\, . 
\end{equation}

Notice that a non-zero Kalb-Ramond 2-form potential leads to a modification of the momentum variable conjugate to $X$ (unless it is pure gauge, in which case the modification is a total time derivative).  The WZ term, which we have neglected so far,  has a similar effect. At the cost of a modification of the constraint functions, one may 
redefine the momentum variables so that $P_m$ and $D^i$  are again canonically conjugate to $X^m$ and $A_i$. For even $p$, and hence a IIA supergravity background,  these modifications are absent in the $T\to 0$ limit that is of most interest to us. However, for odd  $p$, and hence a IIB supergravity background,  there is one  ($p$-dependent) modification that survives the $T\to 0$ limit when the background includes a non-zero axion field. We shall present the details for $p=3$ when we  come to consider the tensionless limit of the D3-brane. We shall also have to examine this issue again when we consider the horizon limit in AdS$_5\times S^5$ because  it  differs from the $T\to0$ limit. 

\subsection{Algebra of constraints}

As the DBI action is manifestly invariant under both worldvolume diffeomorphisms and abelian gauge transformation of the BI gauge potential, one might expect 
the constraint functions of the phase-space action to generate the corresponding transformations of the canonical variables. However, one often finds 
that the PB algebra of constraints is only a subalgebra of the original Lie algebra of diffeomorphisms (e.g. Diff$_1 \oplus\ $Diff$_1 \, \subset$ Diff$_2$ for the 
Nambu-Goto string \cite{Brink:1988nh}) or not even a Lie algebra (as for the Dirac membrane \cite{Henneaux:1985kr}). This is possible because for certain actions (which include
those of $p$-brane type) there are gauge transformations that become trivial in phase space because they vanish `on-shell' (i.e. on 
solutions of the equations of motion). All that one can expect {\it a priori}  is that the constraints are first-class; i.e. that their PB algebra is closed in the sense that the PB of any two constraint functions vanishes on the surface defined by the set of all of them. A check of this is therefore a useful check of gauge invariance, and the detailed 
result would be needed for a BRST quantization \cite{Henneaux:1983um}.  

Here we shall restrict to a Minkowski background with zero dilaton and zero form fields, both Kalb-Ramond and  R-R background fields. In this case the constraint 
functions ${\cal H}_\mu$ ($\mu=0,i$) are as in (\ref{hamcons}) but with 
\begin{equation}
\bE_i = \ell^2 E_i\, , \qquad \bB_{ij} = \ell^2 B_{ij} \, , \qquad \bD^i = \ell^{-2}D^i\, , \qquad \bT=T\, . 
\end{equation}
The canonical equal-time PB relations are
\begin{equation}
\left\{X^m(\bfs),P_n(\bfs')\right\}_{PB} = \delta^m_n \delta^p(\bfs-\bfs') \, ,  \quad  
\left\{A_i(\bfs),D^j(\bfs')\right\}_{PB} = -\delta_i^j \delta^p(\bfs-\bfs') \, . 
\end{equation}
The aim now is to find the PB relations of the constraints, thereby generalizing the result of Henneaux  for the Dirac p-brane \cite{Henneaux:1985kr}.  It is convenient to proceed by first defining the functional
\begin{equation}
{\cal H}[ \alpha] = \int\! d^p\! \sigma \, \alpha^\mu(\bfs) {\cal H}_\mu(\bfs) \qquad (\mu=0, i), 
\end{equation}
where the $\alpha^\mu(\bfs)$ are arbitrary smooth functions with support that allows us to integrate by parts without surface terms. 
Notice that this ${\cal H}[\alpha]$ is invariant under the abelian gauge transformation generated by the constraint function ${\cal J}$, with which 
it therefore has a zero Poisson Bracket. To determine the remaining PB relations it is simplest to first verify that 
\begin{eqnarray}
\left\{ X, {\cal H}[ \alpha]\right\}_{PB} &=& \alpha^0P + \alpha^i\partial_i X \nonumber \\
\left\{P, {\cal H}[ \alpha]\right\}_{PB} &=& \partial_i\left(\alpha^i P\right) + \partial_j \left\{\alpha^0\left[T^2\det (\bK+ \ell^2 B)[\tilde \bK^{-1}]^{ij} + \ell^{-4} D^iD^j\right]\partial_i X\right\} \nonumber \\
\left\{A_i, {\cal H}[ \alpha]\right\}_{PB} &=& -\alpha^0 \ell^{-4} \bK_{ij} D^j - B_{ij} \alpha^j\, , \nonumber \\
\left\{D^i, {\cal H}[ \alpha]\right\}_{PB} &=& \partial_j \left\{ T^2\ell^4 \alpha^0 \det(\bK+ \ell^2 B) \left[ \bK^{-1}B\tilde \bK^{-1}\right]^{ij} + 2D^{[i}\alpha^{j]} \right\}\, . 
\end{eqnarray}
Using these results we find that 
\begin{eqnarray}\label{prev}
\left\{ {\cal H}_i, {\cal H}[\alpha] \right\}_{PB} &=& \alpha^0\partial_i{\cal H}_0 + 2(\partial_i\alpha^0) {\cal H}_0 +
 \alpha^k \partial_k {\cal H}_i + (\partial_i\alpha^k) {\cal H}_k + (\partial_k\alpha^k) {\cal H}_i \nonumber \\
&& +\  \left[\ell^{-4}  \alpha^0 K_{ij} D^j + \alpha^jB_{ij}\right] {\cal J} \, , \nonumber\\
\left\{ {\cal H}_0, {\cal H}[\alpha] \right\}_{PB} &=& \alpha^0\partial_j {\cal O}^i  +  2 (\partial_j\alpha^0){\cal O}^i 
+ \alpha^k \partial_k {\cal H}_0 +  2(\partial_k \alpha^k) {\cal H}_0  \nonumber \\
&& - \  \ell^{-4} \alpha^k (\bK_{kj}D^j) {\cal J} \, , 
\end{eqnarray}
where 
\begin{equation}
{\cal O}^i = \left[ T^2 \det(\bK+ \ell^2 B) [\tilde \bK^{-1}]^{ij} + \ell^{-4}  D^iD^j\right] {\cal H}_j\, . 
\end{equation}

These results, which already show that the set of constraints is first class, imply the following PB relations:
\begin{eqnarray}\label{Dp}
\left\{ {\cal H}_0(\bfs) , {\cal H}_0(\bfs')\right\}_{PB} &=& \left[{\cal O}^i(\bfs) + {\cal O}^i(\bfs')\right] \partial_i \,\delta^p(\bfs-\bfs')  \nonumber \\
\left\{ {\cal H}_0(\bfs) , {\cal H}_i(\bfs')\right\}_{PB} &=& \left[{\cal H}_0(\bfs)  + {\cal H}_0(\bfs') \right] \partial_i\, \delta^p(\bfs-\bfs')
- \ell^{-4}\bK_{ij}D^j {\cal J}  \delta^p(\bfs-\bfs')  \nonumber\\
\left\{ {\cal H}_i(\bfs) , {\cal H}_j(\bfs')\right\}_{PB} &=& {\cal H}_i(\bfs) \partial_j\, \delta^p(\bfs-\bfs') 
+ {\cal H}_j(\bfs') \partial_i\,\delta^p(\bfs-\bfs') -  2\partial_{[i}{\cal H}_{j]}  \delta^p(\bfs-\bfs') \nonumber \\
&& + B_{ij} {\cal J} \delta^p(\bfs-\bfs') \,. 
\end{eqnarray}
The algebra of constraints for the standard Dirac p-brane can be read off  from (\ref{Dp}) by ignoring all BI variables; the result
agrees with \cite{Henneaux:1985kr} except for the $\partial_{[i}{\cal H}_{j]}$ term in penultimate line (which 
is identically zero for $p=1$). 

\subsubsection{Tensionless limit}

In string theory, both $T/\hbar$ and $\ell$ are dimensionless constants times a power of the inverse string tension $\alpha'$.  In units for which 
$\hbar=1$, this implies that 
\begin{equation}
\ell^{-2}  \propto T^{\frac{2}{p+1}}, 
\end{equation}
which means that the phase-space constraints have a limit as $T\to0$:
\begin{equation}
{\cal H}_0 = \frac12 g^{mn} P_m P_n \, , \qquad {\cal H}_i = \partial_i X^m P_m - B_{ij} D^j\, , \qquad {\cal J} =\partial_i D^i\, . 
\end{equation}  
We shall give details of this limit for $p=3$ in the following section. Here we observe that the only background field appearing in the constraint functions is 
the spacetime metric, which makes it easy to compute the algebra of constraints for a tensionless Dp-brane in any background. The  {\it non-zero}
PBs are 
\begin{eqnarray}\label{Dp}
\left\{ {\cal H}_0(\bfs) , {\cal H}_i(\bfs')\right\}_{PB} &=& \left[{\cal H}_0(\bfs)  + {\cal H}_0(\bfs') \right] \partial_i\, \delta^p(\bfs-\bfs') \nonumber\\
\left\{ {\cal H}_i(\bfs) , {\cal H}_j(\bfs')\right\}_{PB} &=& {\cal H}_i(\bfs) \partial_j\, \delta^p(\bfs-\bfs') 
+ {\cal H}_j(\bfs') \partial_i\,\delta^p(\bfs-\bfs') - 2\partial_{[i}{\cal H}_{j]}   \delta^p(\bfs-\bfs') \nonumber \\
&& +  B_{ij} {\cal J}  \delta^p(\bfs-\bfs') \,. 
\end{eqnarray}
The only field dependent term is now the one involving $B$, and this term suggests that we consider
a new basis for the constraint functions in which ${\cal H}_i$ is replaced by 
\begin{equation}
\tilde {\cal H}_i = {\cal H}_i + A_i {\cal J}\, . 
\end{equation}
In terms of this new basis,  the non-zero PBs are
\begin{eqnarray}\label{newbasis}
\left\{ {\cal H}_0(\bfs) , \tilde {\cal H}_i(\bfs')\right\}_{PB} &=& \left[{\cal H}_0(\bfs)  + {\cal H}_0(\bfs') \right] \partial_i\, \delta^p(\bfs-\bfs') \nonumber\\
\left\{ {\cal J}(\bfs) , \tilde {\cal H}_i(\bfs') \right\}_{PB}  &=&  {\cal J}(\bfs') \partial_i \delta^p(\bfs-\bfs')\\
\left\{ \tilde {\cal H}_i(\bfs) , \tilde {\cal H}_j(\bfs')\right\}_{PB} &=& \tilde{\cal H}_i(\bfs) \partial_j\, \delta^p(\bfs-\bfs') 
+ \tilde{\cal H}_j(\bfs') \partial_i\,\delta^p(\bfs-\bfs') - 2\partial_{[i}\tilde {\cal H}_{j]}   \delta^p(\bfs-\bfs') \, .  \nonumber 
\end{eqnarray}
These define a Lie algebra\footnote{The new basis also simplifies the algebra of constraints for non-zero tension, but the PB algebra of the ${\cal H}_0$
constraint functions still involves field-dependent coefficients for $p>1$.}. It is the semi-direct sum of the algebra of p-dimensional diffeomorphisms with representations that are
scalar densities of weights $1$ (${\cal J}$)  and $2$ (${\cal H}_0$). 

\section{The D3-brane and its tensionless limit}

We now focus on the $p=3$ case, for which $\ell^{-4}\propto T$. As  the dimensionless constant of proportionality is rescaled by a rescaling of
the BI gauge potential,  we may choose
\begin{equation}
\ell^2 = 1/\sqrt{T}\, ,  
\end{equation}
as has been done in (\ref{BIaction}). 
In this case, the phase-space action takes form 
\begin{equation}
S= \int\! dt\int \! d^3\sigma \left\{ \dot X^m P_m + \bE_i\bD^i - e{\cal H}_0 -u^i {\cal H}_i\right\} + S_{WZ} \, , 
\end{equation}
with 
\begin{eqnarray}
{\cal H}_0 &=& \frac12\left[ P^2 +  \left( \bD^i \bD^j + \bT^2 \bB^i\bB^j \right)\bK_{ij} + \bT^2\det \bK\right] \nonumber\\
{\cal H}_i &=& \partial_i X^m P_m - \varepsilon_{ijk} \bD^j\bB^k\, , 
\end{eqnarray}
where
\begin{equation}
\bB^i =\frac12\varepsilon^{ijk} \bB_{jk}\, . 
\end{equation}
Since $\ell^{-4}=T$, we now have
\begin{eqnarray}
\bE_i &=& \frac{1}{\sqrt{T}} E_i \, , \qquad \bD^i =  \sqrt{T}\,  D^i \, , \nonumber\\
\bB^i &=&  \frac{1}{\sqrt{T}}\hat B^i \, , \qquad  \left(\hat B^i  = B^i - \frac{\sqrt{T}}{2}\varepsilon^{ijk} \bC_{jk}\right) \, . 
\end{eqnarray}
Recalling that $\bT= Te^{-\phi}$, this yields 
\begin{eqnarray}\label{Hs}
{\cal H}_0 &=& \frac12\left[ P^2 + \bT \left( e^\phi D^iD^j + e^{-\phi} \hat B^i\hat B^j \right) \bK_{ij} + \bT^2 \det\bK\right]\nonumber \\
{\cal H}_i &=& \partial_i X^m P_m - \varepsilon_{ijk}D^j\hat B^k
\end{eqnarray}

Taking the $T\to 0$ limit we arrive at the action 
\begin{equation}\label{tensionless}
S= \int\! dt\int \! d^3\!\sigma \left\{ \dot X^m P_m + E_iD^i - e{\cal H}_0 -u^i {\cal H}_i\right\} + \lim_{T\to0} S_{WZ} \, , 
\end{equation}
where
\begin{equation}
{\cal H}_0  = \frac12 P^2  \, , \qquad {\cal H}_i =  \partial_i X^m P_m + \varepsilon_{ijk}D^j B^k\, , 
\end{equation}
but we must now consider the WZ term; this takes the form
\begin{equation}
S_{WZ} = T \int \left[ \bA_4 + \bA_2 \wedge \bF   + \chi \bF\wedge \bF\right]
\end{equation}
where the integral is over the 4D worldvolume and  $\bA_4$ and $\bA_2$ are the worldvolume $4$-form and $2$-form induced from the 
R-R 4-form and 2-form gauge potentials of IIB supergravity, and $\chi$ is the IIB axion field. Since $\ell^2= 1/\sqrt{T}$ we have 
\begin{equation}
\bF = \frac{1}{\sqrt{T}} F \, , 
\end{equation}
and hence 
\begin{equation}
 \lim_{T\to 0} S_{WZ}  = \int  \chi F\wedge F = \int dt \int d^3\sigma \, \chi \, E_i B^i\, . 
\end{equation}
Using this result, the action (\ref{tensionless}) becomes 
\begin{equation}
S= \int\! dt\int \! d^3\!\sigma \left\{ \dot X^m P_m + E_i \hat D^i - e{\cal H}_0 -u^i {\cal H}_i\right\} \, , 
\end{equation}
where
\begin{equation}
\hat D^i = D^i + \chi B^i \, . 
\end{equation}
The constraint functions should now be rewritten in terms of $\hat D^i$ in place of $D^i$, but their form is {\it unchanged} by this 
substitution. We may therefore drop the `hat' on $\hat D^i$.
\bigskip 

{\bf To summarize}: the tensionless limit of the action for a D3-brane in an {\it arbitrary} bosonic IIB supergravity background is 
\begin{equation}\label{Summact1}
S= \int\! dt\int \! d^3\!\sigma \left\{ \dot X^m P_m + E_i D^i - e{\cal H}_0 -u^i {\cal H}_i\right\} \, , 
\end{equation}
with
\begin{equation}\label{zeroTcons}
{\cal H}_0  =  \frac12 g^{mn}P_m P_n   \, , \qquad {\cal H}_i =  \partial_i X^m P_m - \varepsilon_{ijk} D^j B^k\, . 
\end{equation}
The only dependence on the IIB background comes from the (inverse) spacetime metric in ${\cal H}_0$. 
A fact of importance here is that this  is  the {\it Einstein-conformal-frame} metric.
\bigskip

We record here the field equations that follow from the action (\ref{Summact1}):
\begin{eqnarray}\label{eqs}
\dot X^m  &=& u^i \partial_i X^m + eg^{mn} P_n \nonumber \\
\dot P_m &=&   e\Gamma_{mn}{}^p P_pP^n +  \partial_i(u^iP_m)  \nonumber \\
\dot D^i  &=& - 2\partial_j \left( u^{[i}D^{j]}\right)  \quad \&\quad  \partial_i D^i =0 \nonumber \\
E_i &=&   \varepsilon_{ijk}u^j B^k  \quad \Rightarrow\  \dot B^i =- 2\partial_j \left( u^{[i}B^{j]}\right) \, , 
\end{eqnarray} 
where $\Gamma_{mn}{}^p$ is the usual Levi-Civita connection. These equations must be taken together with the constraints
${\cal H}_\mu=0$ ($\mu=0,1,2,3$).  Notice that the field equations for the electromagnetic fields are completely decoupled from the remaining `branewave' equations.  However, 
any solution of them feeds back into the  `branewave' equations via the momentum constraint. 

\subsection{$Sl(2;\bR)$ electromagnetic duality}

The reason that we have considered in detail the D3-brane in an arbitrary bosonic IIB supergravity background is that the coupling of IIB supergravity 
to a probe D3-brane leads to a promotion of the $SO(2)$ electromagnetic duality invariance of the DBI action to an $Sl(2;\bR)$ duality `covariance', in the sense
that any change in the D3-brane action that results from an  $Sl(2;\bR)$  transformation can be compensated by an $Sl(2;\bR)$ transformation of 
the IIB supergravity background   \cite{Tseytlin:1996it,Green:1996qg}.  What makes this possible is that IIB supergravity has a non-linearly realized $Sl(2;\bR)$ duality 
invariance, with the dilaton and axion parametrizing the coset space $Sl(2;\bR)/SO(2)$ \cite{Schwarz:1983wa}. The only $Sl(2;\bR)$ invariant field 
is the Einstein-conformal frame spacetime metric.

As we have seen, the only background field that is relevant to the tensionless D3-brane is the spacetime metric, in Einstein conformal frame. 
As this is $Sl(2;\bR)$ invariant, the compensating  $Sl(2;\bR)$ duality transformation of the IIB background that is needed for $Sl(2;\bR)$ `covariance'
of the D3-brane for non-zero tension has  no effect on the tensionless D3-brane. It follows that the tensionless D3-brane must have an  
$Sl(2;\bR)$ electromagnetic duality {\it invariance}, in any fixed background. Indeed,  the transformation 
\begin{equation}\label{DBduality}
\left(\begin{array}{c} {\bf D} \\ {\bf B}\end{array}\right) \to S \left(\begin{array}{c} {\bf D} \\ {\bf B}\end{array}\right)  \qquad [S\in Sl(2;\bR] 
\end{equation}
leaves invariant the constraint functions ${\cal H}_\mu$ of (\ref{zeroTcons}), and ${\cal J}$. However it is not defined (at least not locally) as a  transformation 
of  the canonical variables $({\bf D},{\bf A})$, so it is not obviously a symmetry of the action (\ref{Summact1}). 
We now show how this  difficulty can be circumvented. 

\subsubsection{Alternative action with manifest $Sl(2;\bR)$ symmetry}

Recall that 
\begin{equation}
E_i D^i = -\dot A_i D^i - A_0 (\partial_i D^i)  + \partial_i (A_0 D^i) \, . 
\end{equation}
We may omit the total derivative term and then solve the constraint imposed by the Lagrange multiplier $A_0$:
\begin{equation}
D^i = \varepsilon^{ijk} \partial_j \tilde A_k\, . 
\end{equation}
This leads, after integration by parts, to the action\footnote{There may be a relation here to the construction of  \cite{Cederwall:1997ab} in which an action was found that makes manifest the  $Sl(2;\bR)$ covariance of the action for a D3-brane of non-zero tension.}
\begin{equation}\label{Summact}
S= \int\! dt\int \! d^3\sigma \left\{ \dot X^m P_m - \varepsilon^{ijk} \dot A_i \, \partial_j \tilde A_k  - e{\cal H}_0 -u^i {\cal H}_i\right\} \, . 
\end{equation}
This action is invariant under the following 
$Sl(2;\bR)$ transformation of the independent variables, which are now $A_i$ and $\tilde A_i$: 
\begin{equation}
\left(\begin{array}{c} \tilde A_i \\  A_i \end{array}\right) \to S \left(\begin{array}{c} \tilde A_i \\  A_i \end{array}\right)  \qquad [S\in Sl(2;\bR] \, . 
\end{equation}
This $Sl(2;\bR)$ duality transformation implies that of (\ref{DBduality}) and is therefore an invariance of the constraints, but it is also an invariance 
(neglecting  a surface term and a total time derivative) of the mixed `kinetic' term for $(A_i,\tilde A_i)$.  It is therefore an invariance of the action
with corresponding $Sl(2;\bR)$  Noether charges, which are
\begin{equation}
Q_3 = \frac12\int d^3\sigma \left(A_iD^i + \tilde A_i B^i\right)
\end{equation}
and 
\begin{equation}
Q_+ = \int d^3\sigma \, A_iB^i\, , \qquad Q_- = \int d^3\sigma\, \tilde A_i D^i \, . 
\end{equation}

\subsection{Conformal invariance}

As for any tensionless brane, the tensionless D3-brane has an action that is invariant under {\it conformal} isometries of the background 
spacetime metric $g$. To see this, consider a general infinitesimal variation of the type
\begin{equation}
\delta_k X^m=   k^m\, , \qquad \delta_k P_m = - (\partial_m k^n) P_n\, , 
\end{equation}
where $k^m(X)$ is a conformal Killing vector field; i.e. 
\begin{equation}
\left({\cal L}_k g\right)_{mn}   = f g_{mn}
\end{equation}
for some scalar function $f$, where ${\cal L}_k$ is the Lie derivative with respect to $k$.
One finds that the variation of the action (\ref{Summact1}) is 
\begin{equation}
\delta_k S = \frac12 \int\! dt \int \! d^3\!\sigma\, e\,  P^m P^n \left({\cal L}_k g\right)_{mn} =  \int\! dt \int \! d^3\!\sigma\, e f {\cal H}_0\, , 
\end{equation}
which is cancelled if we take $\delta_k e = ef$.  There is therefore a symmetry for each conformal Killing vector field $k$ of the background spacetime.

For example, if we choose the AdS$_5\times S^5$ vacuum, the conformal isometries are, at least locally, those of the 10D Minkowski vacuum
because the AdS$_5\times S^5$ metric is conformally flat. This conformal invariance of the tensionless D3-brane in AdS$_5\times S^5$  has
little connection to the (spontaneously broken) 4D conformal invariance of a D3-brane of non-zero tension in AdS$_5$, which arises from a 
re-interpretation of the AdS$_5$  isometry group as  the conformal isometry group of 4D Minkowski space. Even in the UV or IR limits that 
move the D3-brane to, respectively,  the AdS$_5$ boundary or its Killing horizon, one expects to restore only a 4D conformal invariance, 
not a 5D conformal invariance (corresponding to conformal isometries of AdS$_5$) and certainly not a 10D conformal invariance.  

We shall return to this issue later when we consider in some detail, following \cite{Claus:1998mw}, the D3-brane in AdS$_5\times S^5$. First we 
consider in more detail the tensionless D3-brane in the the 10D Minkowski background.

\subsection{Monge gauge and the strong-field limit}

Returning to (\ref{Summact}) we now choose a Minkowski background with standard Minkowski coordinates, so $g_{mn} = \eta_{mn}$. We may fix the
 reparametrization invariance in this background by choosing the Monge gauge condition
\begin{equation}
X^m(\xi) = \delta^m_\mu \xi^\mu \qquad m=0,1, \dots, 3\, . 
\end{equation}
Let $\vec X(\xi)$ denote the remaining $6$ space coordinates. In this gauge the constraints may be solved for $P_\mu= (P_0,{\bf P})$:
\begin{equation}
P^0 =\pm \sqrt{|{\bf P}|^2 + |\vec P|^2}\, , \qquad {\bf P}= - (\bfn \vec X) \cdot \vec P + {\bf D} \times {\bf B}\, . 
\end{equation}
In addition, the Hamiltonian is $P^0$; choosing it to be positive, we find that the action is 
\begin{equation}\label{sixscalars}
S= \int\! dt\int\! d^3\!\sigma \left\{ \dot{\vec X} \cdot \vec P + E_i D^i - \sqrt{ |\vec P|^2 + \left| (\bfn \vec X) \cdot \vec P - {\bf D} \times {\bf B}\right|^2} \right\}
\end{equation}

The field equations of this action can be found directly or simply by substituting the Monge gauge conditions into the covariant field equations
recorded in  (\ref{eqs}), for the special case of $g_{mn}=\eta_{mn}$.  By doing this one finds that 
\begin{equation}\label{Pmu}
\left(P_0,{\bf P}\right)  = - e^{-1} \left( 1, {\bf u}\right) \, ,  
\end{equation}
which determines the Lagrange multipliers in terms of the $4$-momentum density $P_\mu$. One also finds that 
\begin{eqnarray}\label{Monge-eqs}
0&=& \dot{\vec X} - {\bf u} \cdot\bfn \vec X - \vec P/P^0\nonumber \\
0&=& \dot{\vec P}  - \bfn \cdot ({\bf u} \vec P) \nonumber\\
0&=& \dot {\bf D}  + \bfn \times  ({\bf u} \times {\bf D})  \nonumber\\
0&=& \dot {\bf B}  + \bfn \times ({\bf u} \times {\bf B} ) \, .
\end{eqnarray}

\section{Bialynicki-Birula electrodynamics}

Let us consider solutions of  (\ref{Monge-eqs}) for which  $\vec X$ is a uniform constant $6$-vector; in which case $\vec P \equiv \vec 0$
and we have 
\begin{equation}\label{HP}
P^0 = \left|{\bf D}\times {\bf B}\right| \, , \qquad {\bf P} =  {\bf D}\times {\bf B}\, , 
\end{equation}
so it follows from (\ref{Pmu}) that 
\begin{equation}
-{\bf u} = {\bf P}/P^0  = \frac{{\bf D}\times {\bf B}}{|{\bf D}\times {\bf B}|} \equiv {\bf n}\, . 
\end{equation}
Notice that ${\bf n}$ is a {\it unit} $3$-vector. The equations of motion are now 
\begin{equation}\label{BBeofm}
\dot {\bf D}  =   \bfn \times ({\bf n}\times {\bf D} )\, , \qquad \dot {\bf B}  =  \bfn \times ({\bf n}\times {\bf B} )\, . 
\end{equation}
These are the equationsfound in 1983 by Bialynicki-Birula as a strong-field limit of the Born-Infeld equations \cite{BialynickiBirula:1984tx}. 
As we have seen, they follow from the phase-space action
\begin{equation}\label{BBact}
S_{BB}  = \int\! dt\int\! d^3\!\sigma \left\{{\bf E}\cdot {\bf D} - \left|{\bf D}\times {\bf B}\right| \right\}\, ,  
\end{equation}
given that $E= \bfn A_0 - \dot{\bf A}$ and ${\bf B}=\bfn \times {\bf A}$. The canonical variables are therefore $({\bf A}, {\bf D})$, with canonical PB relations
\begin{equation}\label{BBEpbs}
\left\{A_i(\bfs), D^j(\bfs')\right\}_{PB} = -\delta_i^j \, \delta^3(\bfs-\bfs')\, ,  
\end{equation}
while $A_0$ imposes the constraint  $\bfn \cdot {\bf D}=0$.  The field equations are Lorentz invariant, despite appearances, and even conformally invariant, as was observed by
Bialynicki-Birula, who also discovered that there is an $Sl(2;\bR)$ electromagnetic duality invariance. The name Bialynicki-Birula electrodynamics (BBE) for this 4D field theory is 
therefore appropriate; a more recent discussion of it can be found in  \cite{Gibbons:2001gy}.

The above derivation of BBE from the D3-brane suggests that BBE itself may be recast as the dynamics of a ``space-filling'' D3-brane. 
A brane is space-filling if its worldvolume has the same dimension as spacetime; in the current context this is equivalent to a truncation of the D3-brane 
equations in a 10D Minkowski background. Prior to gauge fixing, this yields the action  
\begin{equation}\label{spacefilled}
S= \int\! dt\int\! d^3\!\sigma \left\{\dot X^\mu P_\mu + E_iD^i - e{\cal H}_0 - u^i{\cal H}_i\right\}\, , 
\end{equation}
where 
\begin{equation}
{\cal H}_0 = \frac12 \eta^{\mu\nu} P_\mu P_\nu \, , \qquad {\cal H}_i = \partial_i {X^\mu} P_\mu - \varepsilon_{ijk}D^j B^k\, , 
\end{equation}
and $\mu=0,1,2,3$ {\it only}.  The  Monge gauge $X^\mu(\xi)=\xi^\mu$ now leads directly to the Bialynicki-Birula  action (\ref{BBact}), 
but the diffeomorphism invariant action (\ref{spacefilled}) has some advantages, one of them being its manifest 4D Lorentz invariance.
The Lorentz group acts linearly on  $(X,P)$ and the corresponding Noether charges are 
\begin{equation}\label{Lorentz}
J^{\mu\nu} = 2\int\! d^3\!\sigma \, X^{[\mu}P^{\nu]} \, . 
\end{equation}
It is a trivial matter to verify (i) that these are constants of the motion and (ii) that their PB algebra is the Lorentz algebra.  More generally, there is a symmetry of  the action (\ref{spacefilled}) for every conformal Killing vector field $k$ of the 4-dimensional Minkowski space, and the
corresponding Noether charge is 
\begin{equation}\label{confcharge}
Q[k] = \int \! d^3\sigma \, k^\mu(X) P_\mu\, . 
\end{equation}

 However, there is a puzzle here: the  electromagnetic fields are inert since $J^{\mu\nu}$ does not depend on them. How can this be correct? 
 Should we not have to assume, at the very least, that  $({\bf D},{\bf B})$ transform as $3$-vectors under the rotation subgroup of the Lorentz group that acts on $(X,P)$? 
The answer is no!  It is consistent, even required, that $({\bf D},{\bf B})$ be regarded as triplets of scalar fields in the action (\ref{spacefilled}) because
their Lorentz transformations emerge upon gauge fixing as a consequence of the fact that the Monge gauge 
condition breaks Lorentz invariance (which is an `internal' symmetry of the action (\ref{spacefilled}) under which the coordinates $\xi^\mu= (t, \bfs)$ do not transform). 
This fact means that a compensating gauge transformation is needed to ensure that a Lorentz transformation preserves the gauge condition, and this results in non-trivial 
Lorentz transformations for $({\bf D},{\bf B})$ {\it after} gauge fixing.   It is straightforward,  in principle,  to determine  the new symmetry  transformations 
that respect the gauge choice by taking into account the compensating gauge transformations, but there is a simpler method.  Noether charges are 
unaffected by compensating gauge  transformations because they are gauge invariant, and this allows an easy determination of the Lorentz transformations of 
the fields of the BBE action (\ref{BBact}), as we shall now show in some detail.

\subsection{Lorentz invariance} 

Consider first the total $4$-momentum; prior to gauge fixing this is  ${\cal P}_\mu= \int\! d^3\sigma P_\mu$.  After imposing the Monge gauge we have 
\begin{equation}
{\cal P}^0  = \int\! d^3\!\sigma\, \left|{\bf D}\times {\bf B} \right| \, , \qquad \PP  = \int\! d^3\!\sigma \left({\bf D}\times {\bf B}\right) \, . 
\end{equation} 
These are indeed conserved, as one can verify using the fact that the equations  (\ref{BBeofm}) (and the remaining constraint $\bfn \cdot {\bf D}=0$) 
imply that 
\begin{eqnarray}\label{DBlemma}
\partial_t \left|{\bf D}\times {\bf B} \right|  = -\partial_i \left( n^i \left|{\bf D}\times {\bf B} \right| \right) \, , \qquad 
\partial_t \left({\bf D}\times {\bf B} \right)_i =  -\partial_j  \left( n^j \left({\bf D}\times {\bf B} \right)_i \right) \, . 
\end{eqnarray}
Using the PB relations of the gauge-fixed action (\ref{BBact}) we find, for example, that $\{ {\bf D}, {\cal P}_\mu,\}_{PB} = -\partial_\mu {\bf D}$, 
as could be expected. 

Now consider the rotation $3$-vector generator
\begin{equation}
{\bf J}  = \int\! d^3\!\sigma \, {\bf X} \times {\bf P} \, . 
\end{equation}
Prior to gauge fixing this generates rotations of ${\bf X}$ and ${\bf P}$ while ${\bf D}$ and ${\bf B}$  are inert. 
However, after gauge fixing we have 
\begin{equation}
{\bf J}  = \int\! d^3\!\sigma \, \left\{\bfs \times ({\bf D} \times {\bf B})\right\} \, , 
\end{equation}
and  a calculation using the canonical PB relations (\ref{BBEpbs}) shows that this charge generates rotations of ${\bf D}$ and ${\bf B}$. For example, 
\begin{equation}
\left\{{\bf D}(\bfs) , \bfo \cdot {\bf J} \right\}_{PB} = \bfo \times {\bf D} - [(\bfo \times\bfs) \cdot\bfn] {\bf D}
\end{equation}
which is the expected transformation of a $3$-vector  field under an infinitesimal rotation of the space coordinates with axial vector parameter $\bfo$. 

But translations and rotations  are manifest symmetries of the gauge-fixed action. What about Lorentz boosts?  Prior to gauge fixing,  the $3$-vector Noether charge is, 
\begin{equation}\label{Lsprefix}
{\bf L} = \int\! d^3\! \sigma \left\{ X^0{\bf P} - {\bf X} P^0\right\} \, . 
\end{equation}
On imposing the Monge gauge this becomes
\begin{equation}\label{Ls} 
{\bf L}  = t \, \PP -  \int\! d^3\!\sigma \,  \bfs \, |{\bf D}\times {\bf B}| \, . 
\end{equation}
It is easily verified from (\ref{DBlemma}) that ${\bf L}$ is a constant of the motion. One also finds easily from the canonical PB relations (\ref{BBEpbs}) that 
\begin{equation}
\left\{ {\bf A}, {\bf v}\cdot {\bf L} \right\}_{PB} =  {\bf N}\times{\bf B}\, , 
\end{equation}
where ${\bf v}$ is a constant $3$-vector boost parameter, and 
\begin{equation}
{\bf N}= t {\bf v} - ({\bf v}\cdot\bfs) {\bf n} \, . 
\end{equation}

The analogous transformations of $({\bf B},{\bf D})$ are 
\begin{eqnarray}
\left\{ {\bf B},\,  {\bf v}\cdot {\bf L} \right\}_{PB}  &=&  \bfn \times ({\bf N}\times{\bf B})\, , \nonumber \\
\left\{{\bf D},  \, {\bf v}\cdot {\bf L} \right\}_{PB}  &=&  \bfn \times ({\bf N}\times{\bf D})\, . 
\end{eqnarray}
These may not look like Lorentz boosts, but any doubts can be settled by a computation of the PB relations of $({\bf J},{\bf L})$ from the canonical PBs 
of the gauge-fixed action. To perform this calculation it is convenient to first establish the following intermediate result, valid for any constant uniform 3-vector ${\bf v}$:
\begin{equation} 
\left\{|{\bf D}\times {\bf B}|,{\bf v} \cdot {\bf L}  \right\}_{PB} = {\bf v} \cdot ({\bf D}\times {\bf B})  -
\bfn \cdot \left( {\bf N}\,  |{\bf D}\times {\bf B}|  \right)\, . 
\end{equation}
Notice that integration over  $3$-space yields the result 
\begin{equation}\label{Lorentz0}
\left\{{\bf v} \cdot {\bf L},\, {\cal P}^0  \right\}_{PB} = -{\bf v}\cdot \PP \, . 
\end{equation}
This is consistent with the time-independence of ${\bf L}$ because 
\begin{equation}
\frac{d {\bf L}}{dt} = \frac{\partial {\bf L}}{\partial t} + \left\{ {\bf L}, {\cal P}^0 \right\}_{PB} =  \PP - \PP ={\bf 0} \, . 
\end{equation}
Another convenient intermediate result is 
\begin{equation}
\left\{{\bf v} \cdot {\bf L}, \, \PP  \right\}_{PB} = - {\bf v}\,  {\cal P}^0\, . 
\end{equation}
Notice that this, together with  (\ref{Lorentz0}), confirms the Lorentz invariance of ${\cal P}^2$. 

From these intermediate results, it is not difficult to deduce that 
\begin{equation}
\left\{{\bf v} \cdot {\bf L}, \, {\bf w}\cdot {\bf L}   \right\}_{PB} = -({\bf v}\times {\bf w}) \cdot {\bf J}\, , 
\end{equation}
which is equivalent to 
\begin{equation}\label{tocheck}
\{ L^i,L^j\}_{PB} = - \varepsilon^{ijk} J_k\, . 
\end{equation}
This is exactly what one finds {\it prior to gauge fixing} from a calculation using (\ref{Lsprefix}) and the canonical PBs derived from the action (\ref{spacefilled}). 
Of course, this should not be a surprise because  gauge-fixing cannot  alter the algebra of Noether charges. 

Finally, we remark that in Monge gauge the generic Noether charge of (\ref{confcharge}) is
\begin{equation}
Q[k] = \int\! d^3\sigma\, \left\{ - k^0(t,\bfs) |{\bf D}\times {\bf B}| + {\bf k}(t,\bfs) \cdot {\bf D} \times {\bf B}\right\}\, .
\end{equation}
Using (\ref{DBlemma}) we find,  after integrating by parts and ignoring the resulting surface terms, that  
\begin{equation}
\dot Q[k] = \int\! d^3\sigma\, \left\{ -(\dot k^0 - n^i n^j \partial_i k_j )|{\bf D}\times {\bf B}| + (\dot{\bf k} - \bfn k^0) \cdot {\bf D}\times {\bf B} \right\} =0\, . 
\end{equation}
The last equality follows from the fact that any conformal Killing vector field on Minkowski spacetime satisfies
\begin{equation}
\dot k^0 =f_k\, , \qquad \dot{\bf k} = \bfn k^0\, , \qquad \partial_{(i}k_{j)} =\delta_{ij}  f_k
\end{equation}
for some ($k$-dependent) function $f_k$ on spacetime.

\section{The IR limit in AdS$_5\times$S$^5$}

We shall now return to the D3-brane action for non-zero tension but in the AdS$_5\times S^5$ background of IIB supergravity.
We also set to zero the dilaton and axion, and all form fields except the 5-form of the background itself. The metric can be written as 
\begin{equation}
ds^2_{10} = \left(\frac{R}{z}\right)^2 \left[ dx^\mu dx^\nu \eta_{\mu\nu}  + dz^2 \right] + R^2 d\Omega_5^2\, , 
\end{equation}
where $R$ is the magnitude of the AdS radius of curvature and $d\Omega_5^2$ is the line element for the unit 5-sphere.
The AdS$_5$ coordinates are adapted to its foliation by 4D Minkowski `slices', parametrized by the Minkowski coordinates
$\{x^\mu; \mu=0,1,2,3\}$ and $z$ is an inverse radial coordinate: the AdS boundary is at $z=0$ while $z=\infty$ is 
the Killing horizon. 

Let  $\bar g_{IJ}$ be the metric on the unit 5-sphere in angular coordinates $\{\psi^I; I=1,2,3,4,5\}$,  so that  $d\Omega_5^2 = d\psi^I d\psi^J \bar g_{IJ}$. 
The induced 3-metric is then 
\begin{equation}
\bK_{ij} = \left(\frac{R}{z}\right)^2 \left[ \partial_i x^\mu \partial_j x^\nu \eta_{\mu\nu} + \partial_i z\partial_j z  + z^2 \partial_i \psi^I \partial_j\psi^J \bar g_{IJ}  \right]  \, . 
\end{equation}
Using this in (\ref{Hs})  leads to  the constraint functions
\begin{eqnarray}
 \left(\frac{R}{z}\right)^2 {\cal H}_0 \equiv \tilde {\cal H}_0 &=&  \frac12  \left[ \eta^{\mu\nu}p_\mu p_\nu  + p_z^2  
 + \frac{L^2}{z^2} + {\cal O}(TR^4/z^4) \right]\, ,  \nonumber \\
 {\cal H}_i &=& \partial_i x^\mu p_\mu + \partial_i z p_z + \partial_i \psi^I p_I  - \varepsilon_{ijk} D^jB^k \, , 
\end{eqnarray}
where (in the absence of  WZ terms) the momentum density  $p_I$ is canonically conjugate to the angular variable $\psi^I$, and 
$L^2$ is the squared magnitude of the  angular  momentum due to motion on $S^5$; it is a quadratic function of the $p_I$ with $\psi_I$-dependent coefficients. 
Notice that all terms in  $\tilde {\cal H}_0$ involving ${\bf D}$ or ${\bf B}$ appear in the ${\cal O}(TR^4/z^4)$ term.

Since the D3-brane tension is not dimensionless, the limit $T\to 0$ makes physical sense only if one specifies it as
a limit of some dimensionless quantity that depends on $T$. In our earlier analysis of the D3-brane in a Minkowski vacuum, 
it was implicit that this dimensionless quantity is the ratio of $T$ to the BI energy density, so that we were taking $T\to0$ at fixed 
BI energy density. An equivalent limit is to take the BI energy density to infinity at fixed $T$, which is a strong-field limit. 
Now, in the AdS$_5\times S^5$ vacuum,  there is an alternative because $T$ appears in the phase-space constraints (and in the WZ term) 
only through $TR^4$,  which is dimensionless (in units for which $\hbar=1$). We can now take $T\to0$ at fixed $R$ (equivalent to taking 
$R\to0$ at fixed $T$), but the result is essentially the same as before: it leads  to  (\ref{Summact1}) but now with the AdS$_5\times S^5$ metric 
appearing in the Hamiltonian constraint function ${\cal H}_0$.

There is another alternative. Consider the limit $R/z\to 0$, which we can interpret as the limit $z\to\infty$ at fixed $R$. This makes sense
if it is applied to the 4D field theory defined by fluctuations of a static planar D3-brane at fixed inverse-radial coordinate $z$ in AdS$_5$; it can be 
viewed as an IR limit, just as $z\to 0$ is a UV limit. Geometrically, it is the limit in which the D3-brane `vacuum' is the AdS$_5$ Killing horizon. 
The main point of this for us is  that the dimensionless parameter $(R/z)^4$ appears everywhere in the rescaled D3-brane constraint function $\tilde{\cal H}_0$ 
with a factor of $T$, so that the limit $z\to\infty$ eliminates its $T$-dependence.  

However, there is a still a factor $T$ in the WZ term, given 
explicitly in \cite{Claus:1998mw}, and this will lead to $T$-dependent redefinitions of the momentum variables of the phase-space action.
A simplification of the $z\to\infty$ limit is that we can ignore the $L^2/z^2$ term in $\tilde{\cal H}_0$. This has the consequence that the $p_I$ become Lagrange multipliers
for the constraints $\partial_i \psi^I=0$, which  makes the $T$-dependent redefinitions of $p_I$ irrelevant, and  implies that the D3-brane has a fixed position on $S^5$, independent 
of position on the brane. Because the IIB $5$-form field strength is self dual, 
there will also be redefinitions of $p_\mu$ and $p_z$, but these come with a factor of  $(TR^4/z^4)$ and can be ignored in the $z\to\infty$ limit. 

The net result of the $z\to\infty$ limit is the phase-space action 
\begin{equation}
S= \int\! dt\int\! d^3\!\sigma \left\{ \dot x^\mu p_\mu + \dot z\, p_z + E_iD^i  - \tilde e \tilde {\cal H}_0 - u^i{\cal H}_i\right\}\, , 
\end{equation}
where
\begin{equation}
\tilde{\cal H}_0 =  \frac12(\eta^{\mu\nu}p_\mu p_\nu + p_z^2) \, , \qquad {\cal H}_i = \partial_i x^\mu p_\mu + \partial_i z\,  p_z - \varepsilon_{ijk} D^j B^k \, . 
\end{equation}
In the Monge gauge $x^\mu(\xi)= \xi^\mu$ the constraints ${\cal H}_\mu$ can be solved for $p_\mu$:
\begin{equation}
p^0  = \sqrt{ |{\bf p}|^2  + p_z^2} \, , \qquad  {\bf p} = - \bfn\! z\,  p_z + {\bf D}\times {\bf B} \, . 
\end{equation}
The gauge fixed action is then 
\begin{equation}
S = \int\! dt\int\! d^3\!\sigma \left\{ \dot z p_z + E_i D^i - \sqrt{p_z^2 + | \bfn\! z\,  p_z - {\bf D}\times {\bf B}|^2} \right\}\, . 
\end{equation}
This is just (\ref{sixscalars}) with one scalar instead of six; i.e. a generalization of BBE to include a scalar field which represents fluctuations
away from the AdS$_5$ horizon of a D3-brane that is otherwise coincident with it.  If we do not allow fluctuations away from the AdS$_5$ horizon 
then we get a further reduction of the dynamics to BBE.

\section{Discussion}

In our introduction we recalled how the Freund-Rubin vacuum of 11D supergravity played a major role in the circle of ideas associated with singletons, AdS, 
KK theory and membranes. This was a precursor of the AdS/CFT correspondence that can be applied, as for AdS/CFT,   to the other ``AdS$\times S$'' 
vacua of M-theory, and we have returned to some of these old ideas in the context of the D3-brane in AdS$_5\times S^5$, specifically the field theory defined by
fluctuations of an infinite planar D3-brane at a fixed distance from the AdS$_5$ Killing horizon.   It has been long understood that this field theory has
a spontaneously broken conformal invariance that is restored in the UV and IR limits, and that the UV limit yields a free field theory on the AdS boundary, but 
the IR limit has not (to our knowledge)  been investigated previously. 

As we hope to have shown here, this limit is a non-trivial one for the D3-brane in that it leads to an interacting 4D field theory that 
generalizes the strong-field limit of Born-Infeld electrodynamics found by Bialynicki-Birula only a few years after the work 
of Freund-Rubin. Here we have shown how a D3-brane re-interpretation of Bialynicki-Birula electrodynamics (BBE) and its generalizations, makes manifest 
many of the remarkable properties of this class of 4D field theory, in particular the $Sl(2;\bR)$ electromagnetic duality invariance. 

These results required the use of the phase-space formulation of the DBI action. This is a well-developed topic but we have provided some pedagogical 
details here of the bosonic sector of the phase-space action for a Dp-brane, and a new result is the Poisson Bracket algebra of the phase-space constraints. 
Starting from the phase space action for the D3-brane in a particular background one may consider various limits that cannot be taken in the DBI action. 
The simplest of these is the tensionless limit in a Minkowski background, which leads to a multi-scalar generalization of BBE. 

However, a slightly different limit is needed to make contact with the idea of  branes as the substrate for singleton dynamics in AdS. This is the IR (horizon) limit 
mentioned above;  it implies zero tension but it also eliminates any motion in $S^5$.  Now we get a D3-brane that is restricted to AdS$_5$; its single scalar 
represents fluctuations away from the AdS horizon, and  the electromagnetic fields propagate on this fluctuating D3-brane. We now recover BBE
as the theory of a non-fluctuating tensionless D3-brane that is coincident with the AdS$_5$ horizon.

We omit the usual comments on ``how our results may be extended in various directions''. We suspect that it may be more useful to the progress
of physics to mention Peter Freund's wonderful book ``A Passion for Discovery'' \cite{Freund:2007},  which deserves to be better known.  The reader will 
find in it many amusing sketches of famous physicists that Peter knew personally, with new or little-known anecdotes, all part of a narrative journey
through the 20th century world of theoretical physics.  If we, the authors, were to write a similar (time-translated) book, Peter Freund would
figure large in it.

\section*{Acknowledgements}
We thank the organisers and participants of the 2019 Benasque {\sl Workshop on Supergravity and Superstrings}, and the staff of the Pedro Pasquale Center for Science, 
for the stimulating environment  that we enjoyed while working on this paper.  The work of PKT has been partially supported by STFC consolidated grant ST/L000385/1.

\providecommand{\href}[2]{#2}\begingroup\raggedright\endgroup


\begin{thebibliography}{10}

  
    
\bibitem{Freund:1980xh}
  P.~G.~O.~Freund and M.~A.~Rubin,
  ``Dynamics of Dimensional Reduction,''
  Phys.\ Lett.\ B {\bf 97} (1980) 233
  
\bibitem{Dirac:1963ta}
  P.~A.~M.~Dirac,
  ``A Remarkable representation of the 3 + 2 de Sitter group,''
  J.\ Math.\ Phys.\  {\bf 4} (1963) 901.
  
\bibitem{Flato:1980we}
  M.~Flato and C.~Fronsdal,
  ``Quantum Field Theory of Singletons: The Rac,''
  J.\ Math.\ Phys.\  {\bf 22} (1981) 1100.
  
  
\bibitem{Sezgin:1983ik}
  E.~Sezgin,
  ``The Spectrum of the Eleven-dimensional Supergravity Compactified on the Round Seven Sphere,''
  Phys.\ Lett.\  {\bf 138B} (1984) 57.

  
\bibitem{Nicolai:1984gb}
  H.~Nicolai and E.~Sezgin,
  ``Singleton Representations of Osp($N$,4),''
  Phys.\ Lett.\  {\bf 143B} (1984) 389.
  
    
\bibitem{Bergshoeff:1987cm}
  E.~Bergshoeff, E.~Sezgin and P.~K.~Townsend,
  ``Supermembranes and Eleven-Dimensional Supergravity,''
  Phys.\ Lett.\ B {\bf 189} (1987) 75.
  
\bibitem{Bergshoeff:1987dh}
  E.~Bergshoeff, M.~J.~Duff, C.~N.~Pope and E.~Sezgin,
  ``Supersymmetric Supermembrane Vacua and Singletons,''
  Phys.\ Lett.\ B {\bf 199} (1987) 69.
  
\bibitem{Duff:1990xz}
  M.~J.~Duff and K.~S.~Stelle,
  ``Multimembrane solutions of D = 11 supergravity,''
  Phys.\ Lett.\ B {\bf 253} (1991) 113.
  
  
\bibitem{Gibbons:1993sv}
  G.~W.~Gibbons and P.~K.~Townsend,
  ``Vacuum interpolation in supergravity via super p-branes,''
  Phys.\ Rev.\ Lett.\  {\bf 71} (1993) 3754
  [hep-th/9307049].
  

  
\bibitem{Dirac:1962iy}
  P.~A.~M.~Dirac,
  ``An Extensible model of the electron,''
  Proc.\ Roy.\ Soc.\ Lond.\ A {\bf 268} (1962) 57.
  
\bibitem{Claus:1998mw}
  P.~Claus, R.~Kallosh, J.~Kumar, P.~K.~Townsend and A.~Van Proeyen,
  ``Conformal theory of M2, D3, M5 and D1-branes + D5-branes,''
  JHEP {\bf 9806} (1998) 004
  [hep-th/9801206].
  
\bibitem{Maldacena:1997re}
  J.~M.~Maldacena,
  ``The Large N limit of superconformal field theories and supergravity,''
  Int.\ J.\ Theor.\ Phys.\  {\bf 38} (1999) 1113
   [Adv.\ Theor.\ Math.\ Phys.\  {\bf 2} (1998) 231]
  [hep-th/9711200].
  
 
  
\bibitem{Schwarz:1983qr}
  J.~H.~Schwarz,
  ``Covariant Field Equations of Chiral N=2 D=10 Supergravity,''
  Nucl.\ Phys.\ B {\bf 226} (1983) 269.
  
 
\bibitem{Horowitz:1991cd}
  G.~T.~Horowitz and A.~Strominger,
  ``Black strings and P-branes,''
  Nucl.\ Phys.\ B {\bf 360} (1991) 197.
  
\bibitem{Ferrara:1997dh}
  S.~Ferrara and C.~Fronsdal,
  ``Conformal Maxwell theory as a singleton field theory on adS(5), IIB three-branes and duality,''
  Class.\ Quant.\ Grav.\  {\bf 15} (1998) 2153
  [hep-th/9712239].
  
  
 
\bibitem{Fradkin:1985ys}
  E.~S.~Fradkin and A.~A.~Tseytlin,
  ``Quantum String Theory Effective Action,''
  Nucl.\ Phys.\ B {\bf 261} (1985) 1
   Erratum: [Nucl.\ Phys.\ B {\bf 269} (1986) 745].
  
\bibitem{Callan:1988wz}
  C.~G.~Callan, Jr., C.~Lovelace, C.~R.~Nappi and S.~A.~Yost,
  ``Loop Corrections to Superstring Equations of Motion,''
  Nucl.\ Phys.\ B {\bf 308} (1988) 221.

\bibitem{Leigh:1989jq}
  R.~G.~Leigh,
  ``Dirac-Born-Infeld Action from Dirichlet Sigma Model,''
  Mod.\ Phys.\ Lett.\ A {\bf 4} (1989) 2767.
  
\bibitem{Li:1995pq}
  M.~Li,
  ``Boundary states of D-branes and Dy strings,''
  Nucl.\ Phys.\ B {\bf 460} (1996) 351
  [hep-th/9510161].

\bibitem{Douglas:1995bn}
  M.~R.~Douglas,
  ``Branes within branes,''
  NATO Sci.\ Ser.\ C {\bf 520} (1999) 267
  [hep-th/9512077].

  
  
\bibitem{BialynickiBirula:1984tx}
  I.~Bialynicki-Birula,
  ``Nonlinear Electrodynamics: Variations on a theme by Born and Infeld'', 
in {\sl Quantum Theory Of Particles and Fields: birthday volume dedicated to  Jan \L{}opusza\'nski}, eds B. Jancewicz and J.  Lukierski, pp 31-48, (World Scientific, 1983). 

  
\bibitem{Bergshoeff:1998ha}
  E.~Bergshoeff and P.~K.~Townsend,
  ``Super D-branes revisited,''
  Nucl.\ Phys.\ B {\bf 531} (1998) 226
  [hep-th/9804011].
  
 

\bibitem{Gauntlett:1997ss}
  J.~P.~Gauntlett, J.~Gomis and P.~K.~Townsend,
  ``BPS bounds for world volume branes,''
  JHEP {\bf 9801} (1998) 003
  [hep-th/9711205].
  
\bibitem{Lindstrom:1997uj}
  U.~Lindstrom and R.~von Unge,
  ``A Picture of D-branes at strong coupling,''
  Phys.\ Lett.\ B {\bf 403} (1997) 233
  [hep-th/9704051].


  
\bibitem{Brink:1988nh}
  L.~Brink and M.~Henneaux, {\sl Principles Of String Theory},  Plenum, N.Y. (1988). 

\bibitem{Henneaux:1985kr}
  M.~Henneaux,
  ``Hamiltonian Form of the Path Integral for Theories with a Gauge Freedom,''
  Phys.\ Rept.\  {\bf 126} (1985) 1.
  
 
  
\bibitem{Henneaux:1983um}
  M.~Henneaux,
  ``Transition Amplitude In The Quantum Theory Of The Relativistic Membrane,''
  Phys.\ Lett.\  {\bf 120B} (1983) 179.
  
\bibitem{Tseytlin:1996it}
  A.~A.~Tseytlin,
  ``Selfduality of Born-Infeld action and Dirichlet three-brane of type IIB superstring theory,''
  Nucl.\ Phys.\ B {\bf 469} (1996) 51
  [hep-th/9602064].
  
\bibitem{Green:1996qg}
  M.~B.~Green and M.~Gutperle,
  ``Comments on three-branes,''
  Phys.\ Lett.\ B {\bf 377} (1996) 28
  [hep-th/9602077].
  
  
\bibitem{Schwarz:1983wa}
  J.~H.~Schwarz and P.~C.~West,
  ``Symmetries and Transformations of Chiral N=2 D=10 Supergravity,''
  Phys.\ Lett.\  {\bf 126B} (1983) 301.


\bibitem{Cederwall:1997ab}
  M.~Cederwall and A.~Westerberg,
  ``World volume fields, SL(2:Z) and duality: The Type IIB three-brane,''
  JHEP {\bf 9802} (1998) 004
  [hep-th/9710007].
  
\bibitem{Gibbons:2001gy}
  G.~W.~Gibbons,
  ``Aspects of Born-Infeld theory and string / M theory,''
  Rev.\ Mex.\ Fis.\  {\bf 49S1} (2003) 19
   [AIP Conf.\ Proc.\  {\bf 589} (2001) no.1,  324]
  [hep-th/0106059].


  \bibitem{Freund:2007}
  Peter Freund, {\sl A Passion for Discovery}, World Scientific 2007. 
  

   
  
   
 

  
 


 

\end{thebibliography}

\end{document}